\documentclass[conference]{IEEEtran}

\usepackage{eqnarray,amssymb,amsmath,amsthm}
\usepackage{mathrsfs}
\usepackage[utf8]{inputenc}
\usepackage[T1]{fontenc}
\usepackage{graphicx}
\usepackage{epsfig}
\usepackage[english]{babel}
\usepackage{subfigure}
\usepackage{color}
\usepackage{epstopdf}
\usepackage{import}
\usepackage{cases}
\usepackage{mathtools}
\usepackage{color, colortbl}
\usepackage{cite}
\usepackage{algorithm}
\usepackage{algpseudocode}
\usepackage{bbm}

\usepackage{array}

\newtheorem{theorem}{Theorem}



\begin{document}

\title{Latency and Reliability-Aware Task Offloading and  Resource Allocation for Mobile Edge Computing\vspace{-0.3cm}}
\author{\IEEEauthorblockN{Chen-Feng Liu\IEEEauthorrefmark{1}, Mehdi Bennis\IEEEauthorrefmark{1}, 
and H.~Vincent Poor\IEEEauthorrefmark{2}}
 \IEEEauthorblockA{\IEEEauthorrefmark{1}Centre for Wireless Communications, University of Oulu, Finland
 \\\IEEEauthorrefmark{2}Department of Electrical Engineering, Princeton University, NJ, USA
 }
\vspace{-1.2cm}
 }

\maketitle

\begin{abstract}
While mobile edge computing (MEC) alleviates the computation and power limitations of mobile devices, additional latency is incurred when offloading tasks to remote MEC servers. In this work, the power-delay tradeoff in the context of task offloading is studied in a multi-user MEC scenario. In contrast with current system designs relying on average metrics (e.g., the average queue length and average latency), a novel network design is proposed in which latency and reliability constraints are taken into account. This is done by imposing a probabilistic constraint on users' task queue lengths and invoking results from \emph{extreme value theory} to characterize the occurrence of low-probability events in terms of queue length (or queuing delay) violation. The problem is formulated as a computation and transmit power minimization subject to  latency  and reliability constraints, and solved using tools from Lyapunov stochastic optimization. Simulation results  demonstrate the effectiveness of the proposed approach, while examining the power-delay tradeoff and  required computational resources for various computation intensities.
\end{abstract}
\begin{IEEEkeywords}
5G, mobile edge computing, fog networking and computing, ultra-reliable low latency communications (URLLC), extreme value theory. 
\end{IEEEkeywords}
\vspace{-0.6em}
\section{Introduction}\label{Sec: Intro}
Mobile edge/fog computing (MEC)  is an architecture that
distributes computation, communication, control, and storage at the network edge \cite{MungChiangFog,fog_survey,Modammed_EuCNC}.
This work is motivated by recent advances in MEC (or fog computing) and surging traffic demands spurred by online video and Internet-of-things applications, including machine type and mission-critical communication (e.g., augmented/virtual reality (AR/VR) and drones).  
While today's communication networks have been engineered with a focus on boosting network capacity, little attention has been paid to latency and reliability performance. Indeed ultra-reliable and low latency communication (URLLC) is one of the most  important enablers of 5G and is currently receiving significant attention in both academia and industry \cite{URLLC}.

When executed at  mobile devices, the performance and quality of experience of computation-intensive applications  are significantly affected by the devices' limited computation capabilities.
Additionally, intensive computations are energy-consuming which severely shortens the lifetime of  battery-limited  devices.  To address the computation and energy issues, mobile devices can wirelessly offload  their tasks  to proximal MEC servers. On the flip side, offloading  tasks incurs extra latency which cannot be overlooked and should be factored into the system design.

\subsection{Related Work}
The energy-delay tradeoff has received significant attention  in MEC networks \cite{JSAC_queue,TaskSplitting,FogPower}.
Kwak {\it et al.}~studied an energy  minimization problem for local  computation, task offloading, and wireless transmission \cite{JSAC_queue}. Therein,  a dynamic task offloading and resource allocation problem was solved using stochastic optimization. 
Taking into account  energy consumption, the authors in  \cite{TaskSplitting} investigated the energy-delay tradeoff of an MEC system in which the user equipment (UE) is endowed with a multi-core central processing unit (CPU) to compute different applications simultaneously. 
In \cite{JSAC_queue} and \cite{TaskSplitting},  one task type  is offloaded to and computed at the single-core CPU MEC server. 
 Furthermore, assuming a multi-core CPU server to compute different users' offloaded tasks in parallel, Mao {\it et al.}~studied a multi-user task offloading and bandwidth allocation problem, in which the energy-delay tradeoff is investigated using the  Lyapunov framework  \cite{FogPower}.

While considering the task queue length as a delay metric,  the works in \cite{JSAC_queue,TaskSplitting,FogPower} only ensure a finite average queue length as time evolves, i.e.,  mean rate stability \cite{Neely/Stochastic}. Nevertheless,  relying on  mean rate stability is insufficient for URLLC applications.
Additionally,  the authors in \cite{FogPower} assumed that the number of users is smaller than the number of server CPU cores, and hence users' offloaded tasks are computed  immediately and simultaneously.
However, if the number of task-offloading users is large, some offloaded tasks need to wait for the available computational resources, incurring latency. In this case, the waiting time for task computing at the server cannot be ignored and  should be taken into account.

\subsection{Our Contribution}

In this work, resource-limited devices seek to offload their computations to nearby MEC servers with multiple CPU cores, while taking into account devices' and servers' computation capabilities, co-channel interference, queuing latency, and reliability constraints. 
The fundamental problem under consideration  is cast as \emph{a network-wide power minimization problem for task computation and offloading, subject to delay requirements and reliability constraints}.
Enabling URLLC mandates a departure from expected utility-based approaches in which relying on average quantities (e.g., average throughput and mean response time) is no longer an option. 
Therefore, motivated by the aforementioned shortcomings, we explicitly model  and characterize  the higher-order statistics of UEs' task queue lengths and impose a probabilistic constraint on the delay/queue length deviation. Further, the queue length bound deviation is directly related to  the URLLC constraint which we examine through the lens of  {\it extreme value theory} \cite{EVT}. Regarding the offloaded tasks at the servers, we take into account the waiting time for the available computational resources and accordingly impose a probabilistic delay requirement.
 Subsequently, invoking tools from Lyapunov stochastic optimization, we propose a dynamic policy for local task execution, task offloading, and resource allocation subject to latency  and reliability constraints.  
Numerical results highlight the effectiveness of the proposed framework while shedding light onto the power-delay tradeoff, and the scaling in terms of servers' computational resources  according to the computation intensity.

%
%
%

\section{System Model}\label{Sec: system}

\begin{figure}[t]
\centering
\includegraphics[width=1\columnwidth]{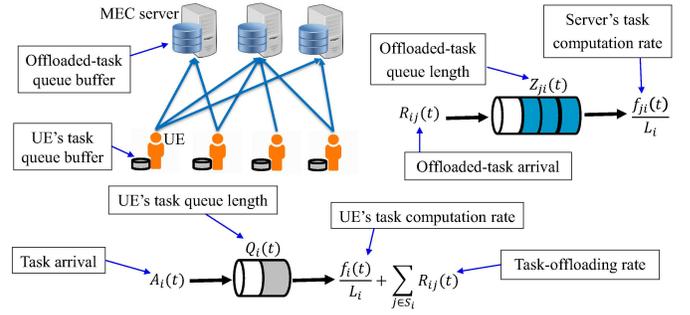}
	\caption{System model.}
\label{Fig: System}
\vspace{-1.5em}
\end{figure}

As shown in Fig.~\ref{Fig: System}, we consider a set $\mathcal{U}$ of  UEs with local computation capability in an MEC network. Due to devices' limited computation capabilities,  a set $\mathcal{S}$ of MEC  servers with  $N$-core CPUs is deployed to help offload and compute UEs' tasks, where each UE wirelessly accesses multiple servers and dynamically offloads its tasks.  The wireless channel gain between UE $i\in\mathcal{U}$ and server $j\in\mathcal{S}$ is denoted by $h_{ij}$ which includes path loss and channel fading.  Based on channel strength, we further assume that each UE $i\in\mathcal{U}$ accesses the set $\mathcal{S}_i$ of servers whose expected channel gains are higher than a threshold value $h_i^{\rm th}$, i.e., $\mathcal{S}_i=\{j\in\mathcal{S}|\,\mathbb{E}[h_{ij}]\geq h_i^{\rm th}\}$. For each server $j\in\mathcal{S}$, the set of accessing UEs is denoted by $\mathcal{U}_j$. Additionally, all channels experience block fading. We index the coherence block by $t\in\{0,1,\cdots\}$, and each block is of unit time length for simplicity.
 
\subsection{Traffic Model at the UE Side}
 Tasks arrive in a stochastic manner following an arbitrary probability distribution. We assume that most tasks' arrivals  can be computed in one time slot while the large-size tasks are divided into small sub-tasks and computed in one time slot \cite{TaskSplitting}. Hereafter, the terminology  ``{\it task}''  refers to both the task which can be computed in one time slot and the divided sub-task.
Each task can be computed at the UE or at the server, following the data-partition model \cite{fog_survey}.
We assume that each offloaded task is assigned to only one server.
Moreover, each UE has a queue buffer to store the tasks' arrivals. 
Denoting the task queue length of UE $i\in\mathcal{U}$ in time slot $t$ as $Q_{i}(t)$,  the queue length (in the unit of bits) evolves as
\begin{align}\label{Eq: local task queue}
Q_i(t+1)&=\max\{Q_i(t)+  A_i(t)-  B_i(t),0\},
\end{align}
where the task arrivals $A_i(t)$ with mean value  $\lambda_i$ are independent and identically distributed over time. Additionally,
\begin{align}\label{Eq: local completion rate}
B_i(t)&=\textstyle\frac{f_i(t)}{L_i}+ \textstyle\sum_{j\in\mathcal{S}_i}R_{ij}(t)
\end{align}
is the task completion rate in time slot $t$ in which 
\begin{equation}\label{Eq: offloading rate}
\textstyle R_{i j}(t)=\frac{W}{|\mathcal{S}|}\log_2\Big(1+\frac{P_{ij}(t)h_{ij}(t)}{\frac{N_0W}{|\mathcal{S}|}+\sum_{i'\in\mathcal{U}_j\setminus i}P_{i'j}(t)h_{i'j}(t)}\Big)
\end{equation}
is the transmission rate\footnote{We assume reception in additive white Gaussian noise.} for task offloading  with transmit power $P_{ij}(t)$ from UE $i\in\mathcal{U}$ to server $j\in\mathcal{S}_i$, while $f_i(t)/L_i$  accounts for the local computation rate of UE $i$ with CPU-cycle frequency $f_i(t)$.
 Here, $L_i$ denotes the required CPU cycles per bit for computation, i.e.,  the processing density, which is application-dependent \cite{JSAC_queue}.
$W$ and $N_0$ are the total system bandwidth and the power spectral density of the additive white Gaussian noise, respectively. 
 Furthermore, the total bandwidth $W$ is orthogonally and equally allocated to different servers whereas UEs share the  allocated band of the accessed server for task offloading. 

We note that given a CPU-cycle frequency $f_i(t)$, the power consumption of local computation is calculated as $\kappa [f_i(t)]^3$ in which the parameter $\kappa$ depends on the hardware architecture \cite{FogPower,CPUhardware}.
In order to minimize local computation power, the UE adopts the dynamic voltage and frequency scaling (DVFS) capability to adaptively adjust its CPU-cycle frequency \cite{fog_survey,CPUhardware}.
Thus, for local computational resource and transmit power allocation, we impose the following constraints for each UE $i\in\mathcal{U}$:
\begin{equation}\label{Eq: local power cost}
\begin{cases}
\textstyle 0\leq f_i(t)\leq f_{i}^{\max},
\\\textstyle\sum_{j\in\mathcal{S}_i}P_{ij}(t)\leq P_{i}^{\max},
\\\textstyle P_{ij}(t)\geq 0,~\forall\,j\in\mathcal{S}_i,
\end{cases}
\end{equation}
where $f_{i}^{\max}$ and $P_{i}^{\max}$ are the maximum local computation capability and transmit power budget, respectively.

\subsection{Traffic Model at the Server Side}
Each MEC server has multiple parallel buffers to process the UEs' offloaded tasks. In this regard, we denote the offloaded-task queue length (in units of bits) for UE $i\in\mathcal{U}_j$ at server $j\in\mathcal{S}$ in time slot $t$ as $Z_{ji}(t)$. The evolution of  $Z_{ji}(t)$ is given by
\begin{align}\label{Eq: computation queue}
Z_{j i}(t+1)&\textstyle\leq \max\big\{Z_{j i}(t)+  R_{ij}(t)-\frac{ f_{ j i}(t)}{L_i},0\big\},
\end{align}
where $f_{ j i}(t)$ is  server $j$'s allocated CPU-cycle frequency for UE $i$'s offloaded tasks.
Since the MEC server  provides a faster computation capability, each CPU core is dedicated to at most one UE (i.e., its offloaded tasks) per time slot. Additionally, we assume that at each server, a given UE's offloaded tasks can only be computed by one CPU core for simplicity.
Thus, each server $j\in\mathcal{S}$ schedules its computation and allocates resources based on the  constraints
\begin{equation}\label{Eq: fog power cost}
\begin{cases}
\sum_{i\in\mathcal{U}_j}\mathbbm{1}\{f_{ji}(t)>0\}\leq N,
\\f_{ji}(t)\in\{0, f_{j}^{\max}\},~\forall\,i\in\mathcal{U}_j,
\end{cases}
\end{equation}
in which $f_{j}^{\max}$ is  server $j$'s computation capability of one CPU-core.

\section{Latency Requirements and Reliability Constraints}\label{Sec: problem}

In this work, the end-to-end delay is composed of the following elements: 1) queuing delay at  the UE and the server, 2) computation delay at the UE and the server, and 3) wireless transmission delay for task offloading.
According to Little's law, the average queuing delay is proportional to the average queue length  \cite{Littlelaw}.
Nevertheless, without considering the probability distribution of the queue length, relying merely on the average queue length fails to account for the low-latency and reliability constraints. To tackle this, we take into account the statistics of the UE's task queue length and impose  a probabilistic constraint
\begin{equation}\label{Eq: ori_net_prob_cst}
\textstyle\lim\limits_{T\to\infty}\frac{1}{T}\sum\limits_{t=1}^{T}\Pr\big(Q_{i}(t)> d_{i} \big)\leq \epsilon_{i},~\forall\,i\in\mathcal{U},
\end{equation}
on the task queue length,
where $d_{i}$ and $\epsilon_{i}\ll1$ are the queue length bound and tolerable violation probability, respectively.
Furthermore, the queue length/queuing delay bound violation also undermines the reliability of task computing. 
For example, if a finite-size queue buffer is over-loaded, the incoming tasks will be dropped.

%
%
\setcounter{equation}{17}
\begin{figure*}[ht]
\begin{align}
Q^{(\rm X)}_{i}(t+1)&\textstyle=\max\Big\{Q^{(\rm X)}_{i}(t)+\Big(X_{i}(t+1)-\frac{\sigma_{i}^{\rm th}}{1-\xi_{i}^{\rm th}}\Big)\times\mathbbm{1}\big\{Q_{i}(t+1)> d_{i}\big\},0\Big\} ,\label{Eq: virtual GPD_user-1}
\\Q^{(\rm Y)}_{i}(t+1)&\textstyle=\max\Big\{Q^{(\rm Y)}_{i}(t)+\Big(Y_{i}(t+1)- \frac{2(\sigma_{i}^{\rm th})^2}{(1-\xi_{i}^{\rm th})(1-2\xi_{i}^{\rm th})}\Big)\times\mathbbm{1}\big\{Q_{i}(t+1)> d_{i}\big\},0\Big\} ,\label{Eq: virtual GPD_user-2}
\\Q^{(\rm X)}_{ji}(t+1)&\textstyle=\max\Big\{Q^{(\rm X)}_{ji}(t)+\Big(X_{ji}(t+1)- \frac{\sigma_{ji}^{\rm th}}{1-\xi_{ji}^{\rm th}}\Big)\times\mathbbm{1}\big\{Z_{ji}(t+1)> \tilde{R}_{ji}(t)d_{ji}\big\},0\Big\} ,\label{Eq: virtual GPD_server-1}
\\Q^{(\rm Y)}_{ji}(t+1)&\textstyle=\max\Big\{Q^{(\rm Y)}_{ji}(t)+\Big(Y_{ji}(t+1)- \frac{2(\sigma_{ji}^{\rm th})^2}{(1-\xi_{ji}^{\rm th})(1-2\xi_{ji}^{\rm th})}\Big)\times\mathbbm{1}\big\{Z_{ji}(t+1)> \tilde{R}_{ji}(t)d_{ji}\big\},0\Big\},\label{Eq: virtual GPD_server-2}
\\Q^{(\rm Q)}_{i}(t+1)&\textstyle=\max\Big\{Q^{(\rm Q)}_{i}(t)
+\mathbbm{1}\big\{Q_{i}(t+1)> d_{i}\big\} -  \epsilon_{i},0\Big\},\label{Eq: physical user}
\\Q^{(\rm Z)}_{ji}(t+1)&\textstyle=\max\Big\{Q^{(\rm Z)}_{ji}(t)
+\mathbbm{1}\big\{Z_{ji}(t+1)>\tilde{R}_{ji}(t) d_{ji}\big\}-  \epsilon_{ji},0\Big\}.\label{Eq: physical server}
\end{align}
\noindent\makebox[\linewidth]{\rule{0.84\paperwidth}{0.4pt}}
\end{figure*}
%
%

Let us further focus on $\bar{F}_{Q_i}(q_i)=\Pr\big(Q_{i}> q_{i} \big)$ in \eqref{Eq: ori_net_prob_cst}, i.e., the complementary cumulative distribution function (CCDF) of the UE's queue length, which reflects the queuing latency profile. If the monotonically decreasing  CCDF decays faster while increasing $q_i$, the probability of having an extreme latency value is lower.
 Since the focus of this work is on extreme case  events which happen  with very low probabilities, i.e., $\Pr\big(Q_{i}> d_{i} \big)\ll 1$, we resort to principles of extreme value theory\footnote{Extreme value theory is a powerful and robust framework to study the tail behavior of a distribution.  Extreme value theory also  provides statistical models for the computation of extreme risk measures \cite{EVT}.} to characterize  the probability distribution of these extreme events and the extreme tails of distributions.
Towards this goal, we first introduce a random variable $Q$ whose cumulative distribution function (CDF) is denoted by $F_{Q}(q)$ and a threshold value $d$. Then, conditioned on $Q>d$, the conditional CDF of the excess value $X=Q-d>0$ is given by 
\begin{align}\notag
\textstyle F_{X|Q>d}(x)=\Pr(Q-d\leq x|Q>d)
=\frac{F_Q(x+d)-F_Q(d)}{1-F_Q(d)}.
\end{align}
\begin{theorem}[{\bf Pickands–Balkema–de Haan theorem for exceedances over thresholds}\cite{EVT}]\label{Thm: Pareto}
Consider the distribution of Q conditioned on exceeding some high threshold $d$. As the threshold $d$ closely approaches $F^{-1}_{Q}(1)$, i.e., $d\to\sup\{q\!\!: F_{Q}(q)<1\}$,  the conditional CDF of the excess value $X=Q-d>0$ is
\begin{equation*}
\textstyle F_{X|Q>d}(x)\approx G(x;\sigma,\xi)=
\begin{cases}
\textstyle 1-\big(\max\big\{1+\frac{\xi x}{\sigma},0\big\}\big)^{-1/\xi},
\\\textstyle\mbox{\hspace{7em}~when~}\xi \neq 0,
\\\textstyle 1-e^{-x/\sigma},\mbox{~when~}\xi =0.
\end{cases}
\end{equation*}
Here, $G(x;\sigma,\xi)$ is the generalized Pareto distribution (GPD) whose  mean and variance  are $\sigma/(1-\xi)$ and $\frac{\sigma^2}{(1-\xi)^2(1-2\xi)}$, respectively.
Moreover, the characteristics of the GPD  depend on the scale parameter $\sigma >0$ and the shape parameter $\xi <1/2$.
\end{theorem}
In other words, Theorem \ref{Thm: Pareto} shows that for sufficiently high threshold $d$, the distribution function of the excess may be approximated by the GPD. 
Now, let us  define the conditional  excess queue value (with respect to the threshold $d_i$ in \eqref{Eq: ori_net_prob_cst}) of each UE $i\in\mathcal{U}$ in time slot $t$ as  $X_i(t)|_{Q_{i}(t)> d_i}=Q_{i}(t)- d_{i}$. 
Then, after applying Theorem \ref{Thm: Pareto},  we obtain the following approximations:
\setcounter{equation}{7}
\begin{align}
\mathbb{E}\big[X_i(t)|Q_{i}(t)> d_{i} \big]&\textstyle\approx\frac{\sigma_i}{1-\xi_i},\label{Eq:remodel_Pareto_mean}
\\\mbox{Var}\big(X_i(t) |Q_{i}(t)> d_{i}\big)&\textstyle\approx\frac{\sigma_i^2}{(1-\xi_i)^2(1-2\xi_i)},\label{Eq:remodel_Pareto_variance}
\end{align}
with a scale parameter $\sigma_i$ and a shape parameter $\xi_i$.
Note that the smaller  $\sigma_i$ and $\xi_i$ are, the smaller are the mean value and variance of the GPD. Hence, we  impose thresholds on the scale parameter and the shape parameter, i.e.,  $\sigma_i\leq \sigma_i^{\rm th}$ and $\xi_i\leq \xi_i^{\rm th}$. Subsequently, applying the two parameter thresholds and $\mbox{Var}(X_i)=\mathbb{E}[X_i^2]-\mathbb{E}[X_i]^2$ to \eqref{Eq:remodel_Pareto_mean} and \eqref{Eq:remodel_Pareto_variance}, we consider the constraints for the time-averaged mean and second moment of the conditional excess queue value, 
\begin{align}
&\hspace{-0.7em}\bar{X}_i\textstyle=\lim\limits_{T\to\infty}\frac{1}{T}\sum\limits_{t=1}^{T}\mathbb{E}\big[X_i(t)|Q_{i}(t)> d_{i} \big]\leq \frac{\sigma_i^{\rm th}}{1-\xi_i^{\rm th}},\label{Eq:GPD-1}
\\&\hspace{-0.7em}\bar{Y}_i\textstyle=\lim\limits_{T\to\infty}\frac{1}{T}\sum\limits_{t=1}^{T}\mathbb{E}\big[Y_i(t) |Q_{i}(t)> d_{i}\big]\leq \frac{2(\sigma_i^{\rm th})^2}{(1-\xi_i^{\rm th})(1-2\xi_i^{\rm th})},\label{Eq:GPD-2}
\end{align}
where $Y_i(t)\coloneqq [X_i(t)]^2$.

Likewise, the average queuing latency at the server is proportional to the ratio of the average queue length to the average offloading rate. Thus, analogous to \eqref{Eq: ori_net_prob_cst}, we consider the constraint
\begin{align}\label{Eq: ori_net_avg_cst}
&\hspace{-0.9em}\textstyle\lim\limits_{T\to\infty}\frac{1}{T}\sum\limits_{t=1}^{T}\Pr\Big(\frac{Z_{ji}(t)}{\tilde{R}_{ij}(t-1)}> d_{ji}\Big)\leq \epsilon_{ji},~\forall\,i\in\mathcal{U}_j,j\in\mathcal{S},
\end{align}
for the offloaded-task queue length, where $\tilde{R}_{ij}(t-1)=\frac{1}{t}\sum_{\tau=0}^{t-1}R_{ij}(\tau)$ is the moving time-averaged offloading rate,
 $d_{ji}$ is the latency bound, and $\epsilon_{ji}\ll 1$ is the tolerable violation probability.
Additionally, for the offloaded-task queue length of UE $i\in\mathcal{U}_j$ at server $j\in\mathcal{S}$,  we define the conditional  exceedance in time slot $t$ as  $X_{ji}(t)|_{Z_{ji}(t)>\tilde{R}_{ij}(t-1) d_{ji}}=Z_{ji}(t)- \tilde{R}_{ij}(t-1)d_{ji}$ with respect to the threshold $\tilde{R}_{ij}(t-1)d_{ji}$.
Similar to the task queue length at the user's side, we have the constraints for the conditional exceedance of the offloaded-task queue length, i.e., $\forall\,i\in\mathcal{U}_j,j\in\mathcal{S}$,
\begin{align}
&\hspace{-0.5em}\bar{X}_{ji}\textstyle=\lim\limits_{T\to\infty}\frac{1}{T}\sum\limits_{t=1}^{T}\mathbb{E}\big[X_{ji}(t)|Z_{ji}(t)>\tilde{R}_{ij}(t-1) d_{ji} \big]\leq \frac{\sigma_{ji}^{\rm th}}{1-\xi_{ji}^{\rm th}},\label{Eq:GPD-3}
\\&\hspace{-0.5em}\bar{Y}_{ji}\textstyle=\lim\limits_{T\to\infty}\frac{1}{T}\sum\limits_{t=1}^{T}\mathbb{E}\big[Y_{ji}(t) |Z_{ji}(t)> \tilde{R}_{ij}(t-1)d_{ji}\big]&\notag
\\&\hspace{13.5em}\textstyle\leq \frac{2(\sigma_{ji}^{\rm th})^2}{(1-\xi_{ji}^{\rm th})(1-2\xi_{ji}^{\rm th})},\label{Eq:GPD-4}
\end{align}
with $Y_{ji}(t)\coloneqq [X_{ji}(t)]^2$ and the  thresholds $ \sigma_{ji}^{\rm th}$ and $\xi_{ji}^{\rm th}$ for the approximated GPD.

At the UE, the computation delay and transmission delay for task offloading are inversely proportional to $f_i(t)$ and $ \sum_{j\in\mathcal{S}_i}R_{ij}(t)$, respectively, as per \eqref{Eq: local completion rate}. To decrease the computation delay, the UE should allocate more local CPU-cycle frequency while computing tasks locally. Analogously, more transmit power is allocated in order to decrease transmission delay while offloading the task.  Since the queue length can also be further reduced by allocating higher CPU-cycle frequency or/and transmit power,  both delays are implicitly taken into account in the queue length constraints \eqref{Eq: ori_net_prob_cst}, \eqref{Eq:GPD-1}, and \eqref{Eq:GPD-2}. On the other hand,  allocating more computation capability or/and transmit power depletes the UE's battery quickly. Therefore, the power-delay tradeoff is crucial and will be investigated in the numerical results section.
Regarding the computation delay at the server, this is negligible since one CPU-core with  better computation capability is dedicated for one UE's offloaded tasks. 
Nevertheless, since queuing delay at the server with the requirements \eqref{Eq: ori_net_avg_cst}-\eqref{Eq:GPD-4}  is considered,
the server needs to schedule computation and allocate CPU cores when the number of  accessing UEs is larger than number of the server's CPU cores.
Taking into account the latency requirements and reliability constraints, we  propose
a dynamic task offloading and resource allocation approach that minimizes the UEs'  power consumption.

\section{Latency and Reliability-aware Task Offloading and Resource Allocation}\label{Sec: approach}

Denoting  the network-wide power allocation vector and the computational resource allocation vector of server $j\in\mathcal{S}$ as $\mathbf{P}(t)=(P_{ij}(t),i\in\mathcal{U},j\in\mathcal{S}_i)$ and $\mathbf{f}_j(t)=(f_{ji}(t),i\in\mathcal{U}_j,j\in\mathcal{S})$, respectively, our studied optimization problem is  formulated as follows:
\begin{IEEEeqnarray}{cl}\label{Eq: ori_net_problem}
\underset{\mathbf{\mathbf{P}}(t),f_i(t),\mathbf{f}_j(t)}{\mbox{minimize}}&~~\textstyle\sum\limits_{i\in\mathcal{U}}\bar{P}_{i}
%
\\\mbox{subject to}&~~\mbox{\eqref{Eq: local power cost} and \eqref{Eq: fog power cost} for resource allocation,}\notag
\\&~~\mbox{\eqref{Eq: ori_net_prob_cst} and \eqref{Eq: ori_net_avg_cst} for delay bound violation,}\notag
\\&~~\mbox{\eqref{Eq:GPD-1},  \eqref{Eq:GPD-2},  \eqref{Eq:GPD-3}, and  \eqref{Eq:GPD-4} for the GPD,}\notag
\end{IEEEeqnarray}
%
%
%
where $\bar{P}_{i}=\lim\limits_{T\to\infty}\frac{1}{T}\sum_{t=0}^{T-1}\big(\kappa [f_i(t)]^3+\sum_{j\in\mathcal{S}_i}P_{ij}(t)\big)$ is  UE $i$'s long-term time average  power consumption.
To find the optimal task offloading and computational resource allocation policy for problem \eqref{Eq: ori_net_problem}, we resort to  techniques from Lyapunov stochastic optimization.

\subsection{Lyapunov Optimization Framework}

We first rewrite \eqref{Eq: ori_net_prob_cst} and \eqref{Eq: ori_net_avg_cst} as
\begin{align}
&\hspace{-0.5em}\textstyle\lim\limits_{T\to\infty}\frac{1}{T}\sum\limits_{t=1}^{T}\mathbb{E}\big[\mathbbm{1}\big\{Q_{i}(t)> d_{i}\big\}\big]\leq \epsilon_{i},\label{Eq:remodel_dev}
\\&\hspace{-0.5em}\textstyle\lim\limits_{T\to\infty}\frac{1}{T}\sum\limits_{t=1}^{T}\mathbb{E}\big[\mathbbm{1}\big\{Z_{ji}(t)> \tilde{R}_{ji}(t-1)d_{ji}\big\}\big]\leq \epsilon_{ji},\label{Eq:remodel_dev-2}
\end{align}
respectively, where $\mathbbm{1}\{\cdot\}$ is  the indicator function.
In order to ensure all time-averaged constraints \eqref{Eq:GPD-1},  \eqref{Eq:GPD-2},  \eqref{Eq:GPD-3}, \eqref{Eq:GPD-4}, \eqref{Eq:remodel_dev}, and \eqref{Eq:remodel_dev-2}, we respectively introduce the corresponding virtual queues \eqref{Eq: virtual GPD_user-1}--\eqref{Eq: physical server} with the dynamics  shown on the top of this page.
Then, letting $\mathbf{Q}_{u}(t)\coloneqq\big(Q^{(\rm X)}_{i}(t),Q^{(\rm Y)}_{i}(t),Q^{(\rm Q)}_{i}(t),i\in\mathcal{U}\big)$ and $\mathbf{Q}_s(t)\coloneqq\big(Q^{(\rm X)}_{ji}(t),Q^{(\rm Y)}_{ji}(t),Q^{(\rm Z)}_{ji}(t),i\in\mathcal{U}_j,j\in\mathcal{S}\big)$ denote the combined queue vectors for notational simplicity, we express the conditional Lyapunov drift-plus-penalty for slot $t$ as
\setcounter{equation}{23}
\begin{equation}\label{Eq: Conditional Lyapunov drift}
\textstyle\mathbb{E}\Big[\mathcal{L}(\mathbf{Q}(t+1))-\mathcal{L}(\mathbf{Q}(t))+\sum\limits_{i\in\mathcal{U}}VP_{i}(t)
\Big|\mathbf{Q}_u(t),\mathbf{Q}_s(t)\Big],
\end{equation}
where $\mathcal{L}(\mathbf{Q}(t))=\frac{1}{2}\sum_{i\in\mathcal{U}}\big[\big(Q^{(\rm X)}_{i}(t)\big)^2+\big(Q^{(\rm Y)}_{i}(t)\big)^2+\big(Q^{(\rm Q)}_{i}(t)\big)^2\big]+\frac{1}{2}\sum_{j\in\mathcal{S}}\sum_{i\in\mathcal{U}_j}\big[\big(Q^{(\rm X)}_{ji}(t)\big)^2+\big(Q^{(\rm Y)}_{ji}(t)\big)^2+\big(Q^{(\rm Z)}_{ji}(t)\big)^2\big]$ is the Lyapunov function. $V\geq 0$ is a parameter that trades off power consumption and end-to-end latency. 
Subsequently, substituting $(\max\{x,0\})^2\leq x^2$ and all physical and virtual queue dynamics into  \eqref{Eq: Conditional Lyapunov drift},  we obtain
\begin{align}
 \eqref{Eq: Conditional Lyapunov drift}&\textstyle\leq C+ \mathbb{E}\Big[-   \sum_{i\in\mathcal{U}}\Big[ B_i(t)\Big(Q^{(\rm X)}_{i}(t)+Q_i(t)+  A_i(t)\notag
\\&\textstyle+2Q^{(\rm Y)}_{i}(t)\big(Q_i(t)+  A_i(t)\big) +2\big(Q_i(t)+  A_i(t)\big)^3\Big)   \notag
\\&\textstyle+Q^{(\rm Q)}_{i}(t)\Big]  \times\mathbbm{1}\big\{\max\{Q_i(t)+  A_i(t)-  B_i(t),0\}> d_{i}\big\}\notag
\\&\textstyle + \sum_{j\in\mathcal{S}} \sum_{i\in\mathcal{U}_j}\Big[\Big(  R_{ij}(t)-\frac{ f_{ j i}(t)}{L_i}\Big)\Big(Q^{(\rm X)}_{ji}(t)+Z_{j i}(t) \notag
\\&\textstyle+2Q^{(\rm Y)}_{ji}(t)Z_{j i}(t)+2[Z_{j i}(t)]^3\Big) +Q^{(\rm Z)}_{ji}(t)  \Big]\notag
\\&\textstyle\times\mathbbm{1}\Big\{\max\Big\{Z_{j i}(t)+  R_{ij}(t)-\frac{ f_{ j i}(t)}{L_i},0\Big\}> \tilde{R}_{ji}(t)d_{ji}\Big\}\notag
\\&\textstyle+\sum_{i\in\mathcal{U}}V\Big(\kappa [f_i(t)]^3+\sum_{j\in\mathcal{S}_i}P_{ij}(t)\Big)\Big|\mathbf{Q}_u(t),\mathbf{Q}_s(t)\Big]. \label{Eq: Lyapunov bound}
\end{align}
Due to space limitations, we omit the details of the constant $C$  which does not affect the system performance. Note that the solution to problem \eqref{Eq: ori_net_problem} can be obtained by minimizing the upper bound in \eqref{Eq: Lyapunov bound}  \cite{Neely/Stochastic}. To this end, we have three optimization problems to be solved in each time slot: two optimization problems for local computation and task offloading at the UE side while the other problem is at the server side.

\subsection{Task Computation and Offloading at the UE Side}

For local computation, UE $i\in\mathcal{U}$ allocates the CPU-cycle frequency by solving
\begin{IEEEeqnarray}{cl}
\textstyle\underset{0\leq f_i(t)\leq f_{i}^{\max}}{\mbox{minimize}}&~~V\kappa [f_i(t)]^3-a_i (t)f_i(t)/L_i\label{Eq: local UE computation problem}
\end{IEEEeqnarray}
with  $a_i(t)=Q^{(\rm Q)}_{i}(t)+Q_i(t)+  A_i(t) +\big[Q^{(\rm X)}_{i}(t)+Q_i(t)+  A_i(t)+2Q^{(\rm Y)}_{i}(t)\left(Q_i(t)+  A_i(t)\right)+ 2\left(Q_i(t)+  A_i(t)\right) ^3\big]\times\mathbbm{1}\{Q_i(t)+  A_i(t)> d_{i}\}$. 
The optimal solution to \eqref{Eq: local UE computation problem}, i.e.,
%
%
%
$\textstyle f_i^{*}(t)=\min\Big\{ \sqrt{\frac{a_i(t) }{3V\kappa L_i}},f_{i}^{\max}\Big\},$
%
%
%
can be obtained by differentiation.

The second  problem for task offloading is given as follows:
\begin{subequations}\label{Eq: UE problem}
\begin{IEEEeqnarray}{cl}
\hspace{-2em}\underset{\mathbf{P}(t)}{\mbox{minimize}}&~~\textstyle
\sum\limits_{i\in\mathcal{U}}\sum\limits_{j\in\mathcal{S}_i}\big[VP_{ij}(t)+\big(b_{j i}(t)-a_i(t)\big)  R_{ij}(t)\big]\label{Eq: UE problem_obj}
\\\hspace{-2em}\mbox{subject to}&~~\textstyle\sum\limits_{j\in\mathcal{S}_i}P_{ij}(t)\leq P_{i}^{\max},~\forall\,i\in\mathcal{U},
\\&~~\textstyle P_{ij}(t)\geq 0,~\forall\,i\in\mathcal{U},j\in\mathcal{S}_i,
\end{IEEEeqnarray}
\end{subequations}
which jointly decides all UEs' task offloading policies.
Here, $b_{ji}(t)=Q^{(\rm Z)}_{ji}(t)+Z_{j i}(t) +\big[Q^{(\rm X)}_{ji}(t)+Z_{j i}(t)+2Q^{(\rm Y)}_{ji}(t)Z_{j i}(t)+2[Z_{j i}(t)]^3\big]\times\mathbbm{1}\big\{Z_{j i}(t)+  R_{i}^{\max}> \tilde{R}_{ji}(t-1)d_{ji}\big\}$, and $R_{i}^{\max}$ is UE $i$'s maximum offloading rate.
Solving \eqref{Eq: UE problem} without a central controller, each UE requires other UEs'  channel state information (CSI) and queue state information (QSI) as per \eqref{Eq: offloading rate} and \eqref{Eq: UE problem_obj}, which is impractical and  incurs high overhead, especially in dense  networks. 
In order to alleviate this problem, we decompose the summation (over all UEs) in the objective \eqref{Eq: UE problem_obj} and reformulate \eqref{Eq: UE problem} as $|\mathcal{U}|$ sub-problems, i.e., $\forall\,i\in\mathcal{U}$,
\begin{subequations}\label{Eq: local UE problem}
\begin{IEEEeqnarray}{cl}
\underset{P_{ij}}{\mbox{minimize}}&~~\textstyle\sum\limits_{j\in\mathcal{S}_i}VP_{ij}+\sum\limits_{j\in\mathcal{S}_i}\big(b_{j i} -a_i\big)\notag
\\&~~\textstyle\times\mathbb{E}_{I_{ij}}\Big[\frac{W}{|\mathcal{S}|}\log_2\Big(1+\frac{|\mathcal{S}|P_{ij}h_{ij}}{N_0W+|\mathcal{S}|I_{ij}}\Big)\Big]
\\\mbox{subject to}&~~\textstyle\sum\limits_{j\in\mathcal{S}_i}P_{ij}\leq P_{i}^{\max},
\\&~~\textstyle P_{ij}\geq 0,~\forall\,j\in\mathcal{S}_i,
\end{IEEEeqnarray}
\end{subequations}
which is solved at UE $i$ in a decentralized way. Here, the expectation is with respect to the estimated distribution of the aggregate interference  $I_{ij}=\sum_{i'\in\mathcal{U}_j\setminus i}P_{i'j}h_{i'j}$, and we suppress the time index $t$ for simplicity. 
In the decomposed problem \eqref{Eq: local UE problem}, each UE decides on the transmit power for task offloading without requiring  other UEs' CSI and QSI. Subsequently, applying
 the Karush-Kuhn-Tucker (KKT) conditions to the convex optimization problem \eqref{Eq: local UE problem}, UE $i$  offloads tasks to server $j$ with the optimal transmit  power $P^{*}_{ij}>0$ satisfying
%
%
$\textstyle\mathbb{E}_{I_{ij}}\big[\frac{(a_i-b_{j i} )Wh_{ij}}{( N_0W+I_{ij}|\mathcal{S}|+P^{*}_{ij}h_{ij}|\mathcal{S}|)\ln2}\big]=V+\gamma$
%
%
%
if $\mathbb{E}_{I_{ij}}\big[\frac{(a_i-b_{j i})Wh_{ij}}{(N_0W+I_{ij}|\mathcal{S}|)\ln2 }\big]>V+\gamma$. Otherwise,  $P^{*}_{ij}=0$. 
The Lagrange multiplier $\gamma$ is 0 if $\sum_{j\in\mathcal{S}_i}P^{*}_{ij}< P_{i}^{\max}$. When $\gamma> 0$,
$\sum_{j\in\mathcal{S}_i}P^{*}_{ij}= P_{i}^{\max}$.

\subsection{Computational Resource Allocation at the Server Side}

At the server side, each MEC server $j\in\mathcal{S}$ allocates computational resources by solving the following  optimization problem:
\begin{subequations}\label{Eq: server problem}
\begin{IEEEeqnarray}{cl}
\underset{f_{ji}(t)}{\mbox{maximize}}&~~\textstyle\sum\limits_{i\in\mathcal{U}_j} b_{j i}(t)f_{ j i}(t)/L_i\label{Eq: server problem-1}
\\\mbox{subject to}&~~\textstyle\sum\limits_{i\in\mathcal{U}_j}\mathbbm{1}\{f_{ji}(t)>0\}\leq N,
\\&~~\textstyle f_{ji}(t)\in\{0, f_{j}^{\max}\},~\forall\,i\in\mathcal{U}_j.
\end{IEEEeqnarray}
\end{subequations}
To maximize \eqref{Eq: server problem-1}, the server equally dedicates its $N$ CPU cores to the UEs with the $N$ largest values of $b_{j i}(t)/L_i$. 
The solution to problem \eqref{Eq: server problem} is detailed in Algorithm \ref{Alg: cloud}.
\begin{algorithm}[t]
  \caption{Computational resource allocation at the server.}
  \begin{algorithmic}[1]
    \State Initialize $n=1$ and $\tilde{\mathcal{U}}=\mathcal{U}_j$.
    \While{$n\leq  N$ and $\tilde{\mathcal{U}}\neq \emptyset$}
      \State $i^*=\operatorname{argmax}_{i\in\tilde{\mathcal{U}}} \big\{b_{j i}(t) /L_i\big\}$.
      \State $f_{ j i^{*}}(t)= f_{j}^{\max}$.
      \State Update $n\leftarrow n+1$ and $\tilde{\mathcal{U}}=\tilde{\mathcal{U}}\setminus i^{*}$.
      \EndWhile
  \end{algorithmic}\label{Alg: cloud}
\end{algorithm}

After computing and offloading the tasks in time slot $t$, the UEs update the queue length in \eqref{Eq: local task queue}, \eqref{Eq: virtual GPD_user-1}, \eqref{Eq: virtual GPD_user-2}, and \eqref{Eq: physical user} while the MEC servers update \eqref{Eq: computation queue}, \eqref{Eq: virtual GPD_server-1}, \eqref{Eq: virtual GPD_server-2}, and \eqref{Eq: physical server}, for the next slot $t+1$. Moreover, based on the transmission rate $R_{ij}(t)$, the UE empirically estimates the statistics of $I_{ij}$ for slot $t+1$ as per $ \hat{\Pr}\big(\tilde{I}_{ij};t+1\big)= \frac{\mathbbm{1}\{I_{ij}(t)=\tilde{I}_{ij}\}}{t+2}+ \frac{(t+1)\hat{\Pr}(\tilde{I}_{ij};t)}{t+2}$.
\begin{figure*}
\centering
\begin{minipage}{0.64\linewidth}
\subfigure[$L=737.5\,$cycle/bit.]{\includegraphics[width=0.5\linewidth]{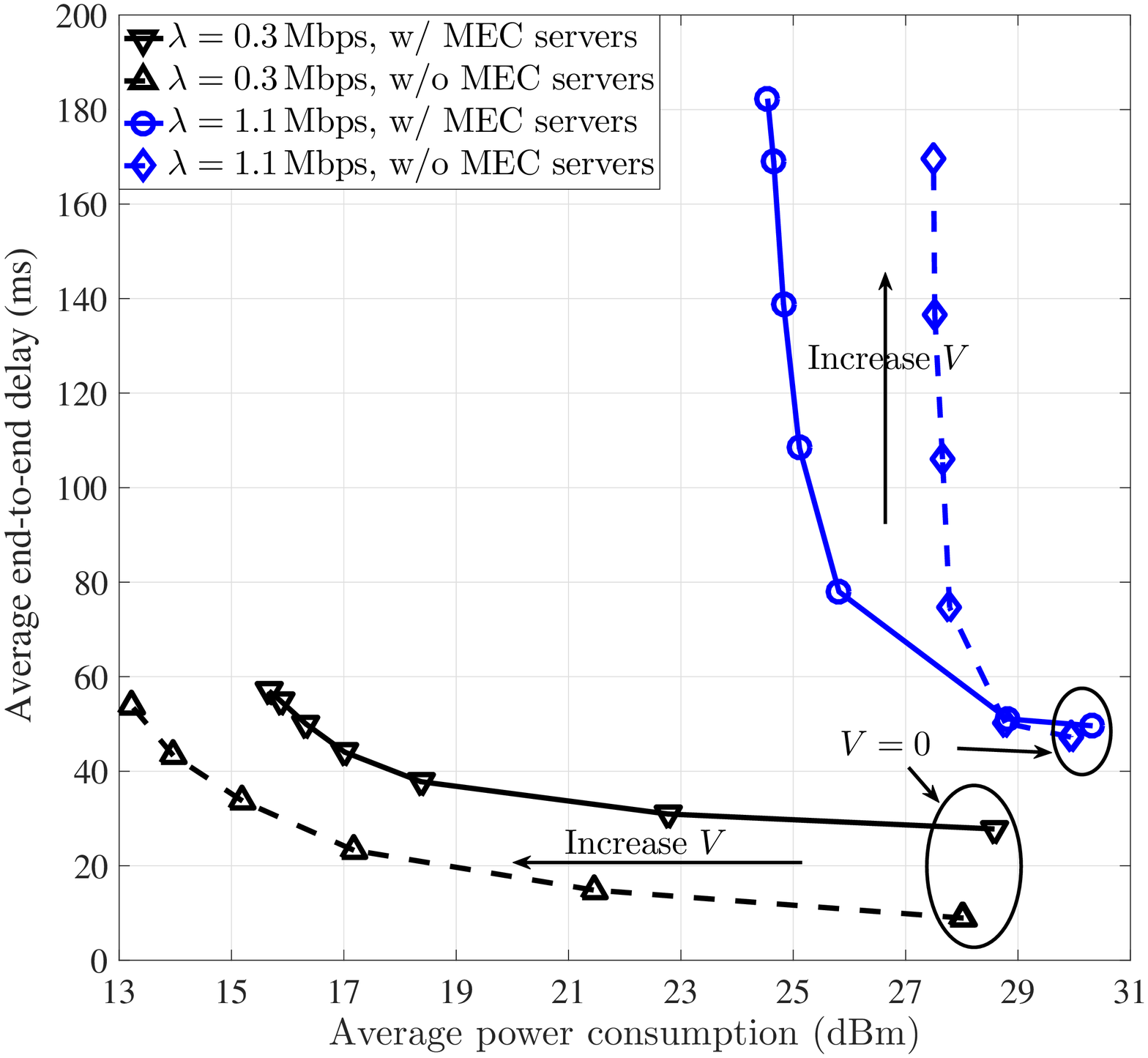}}
\subfigure[ $\lambda=0.3$\,Mbps.]{\includegraphics[width=0.5\linewidth]{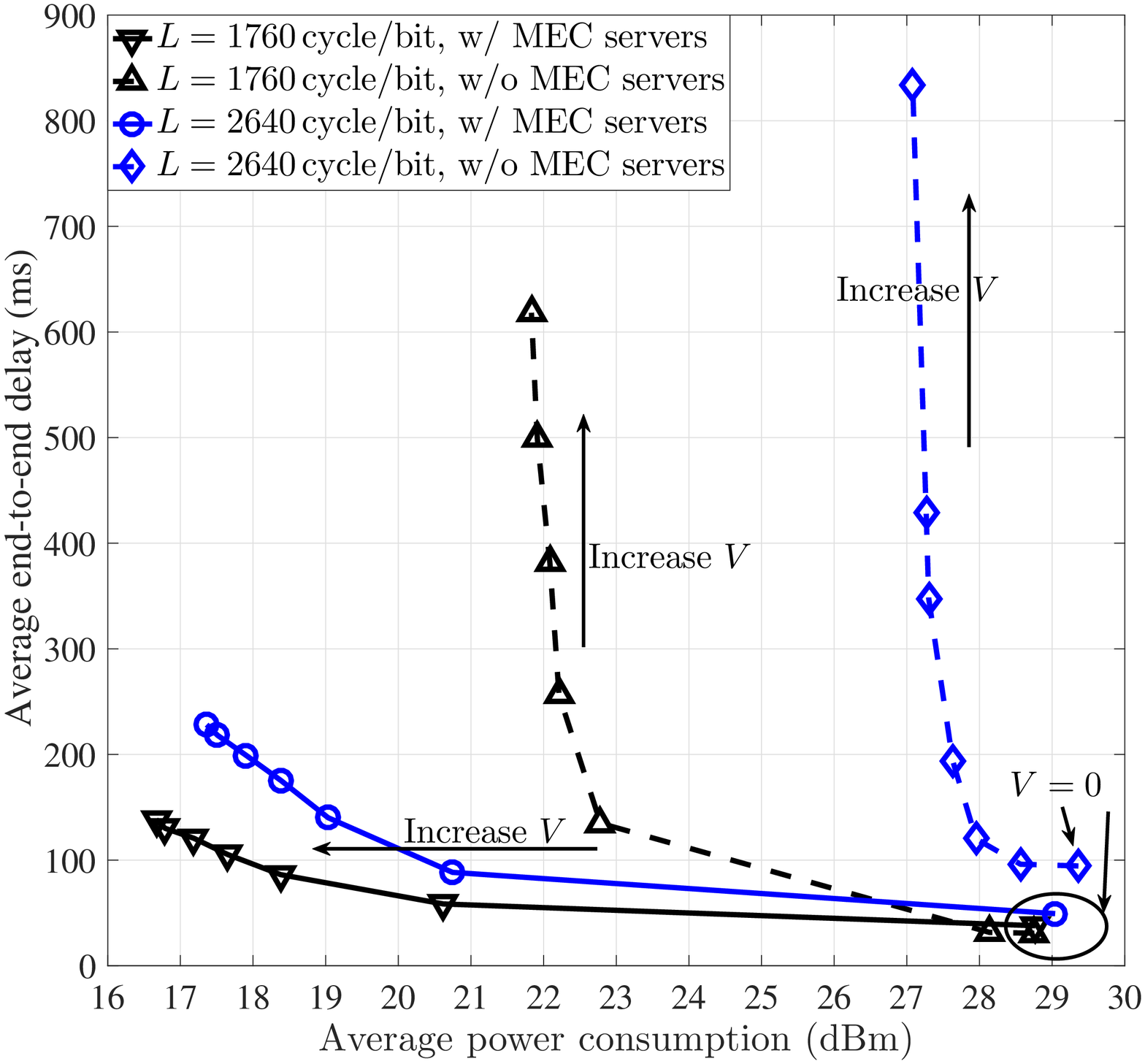}}
	\caption{Tradeoff between  a UE's average power consumption and end-to-end delay.}
	\label{Fig:tradeoff}
\end{minipage}
\hspace{1em}
\begin{minipage}{0.32\linewidth}
\vspace{-0.3em}
	\includegraphics[width=1\columnwidth]{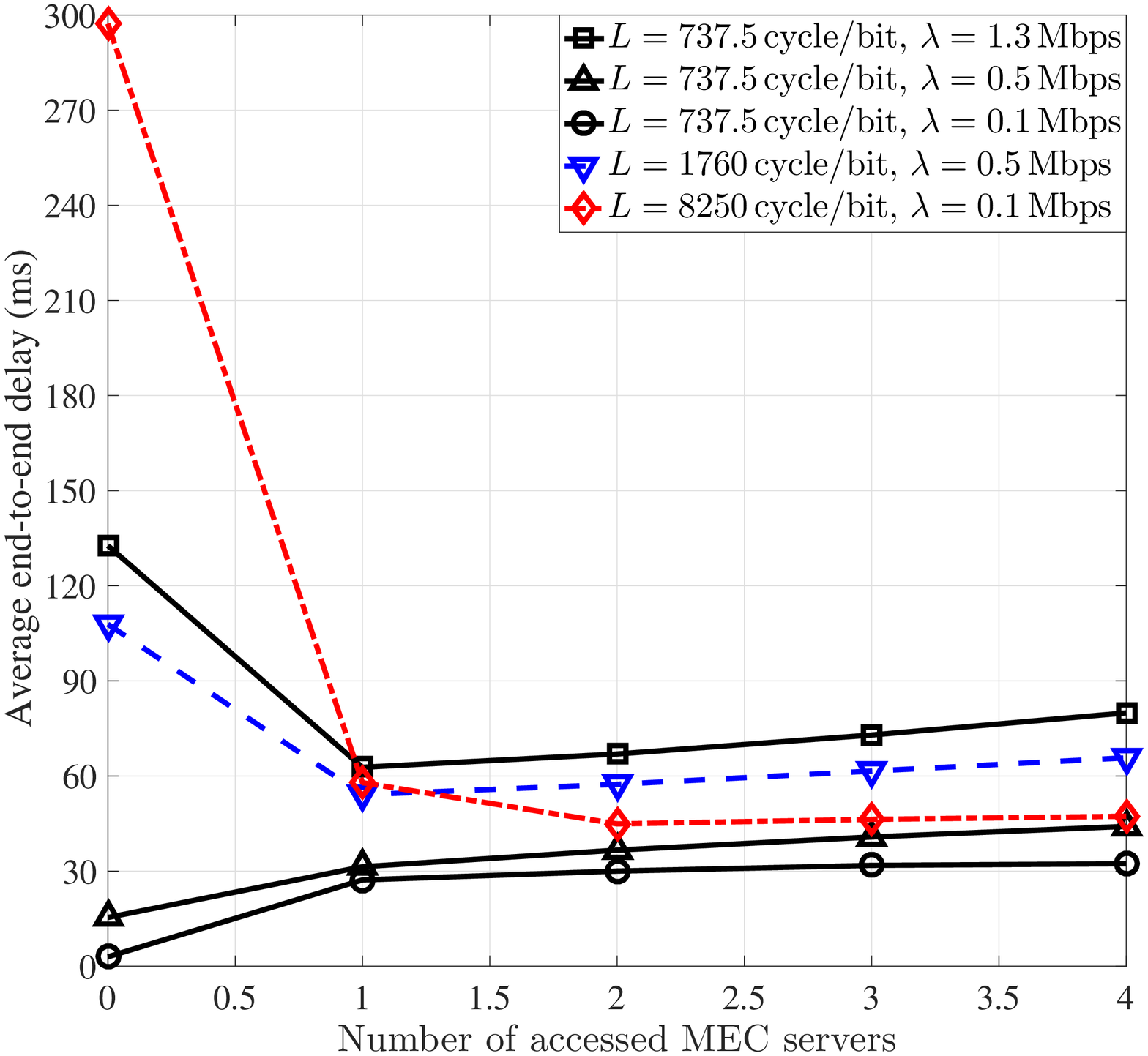}
	\caption{Average end-to-end delay versus number of accessed MEC servers per UE.}
		\label{Fig:accessed number}
\end{minipage}
\vspace{-1.5em}
\end{figure*}

\section{Numerical Results}\label{Sec: results}

\begin{table}[t]
\caption{Simulation parameters \cite{FogPower,Al16,JSAC_queue,Mie2010}.}\label{Tab: parameters}
\centering
 \begin{tabular}{|m{1.2cm}|m{3.1cm}|m{1.2cm}|m{1.3cm}|}
\hline
{\bf Parameter} &{\bf  Value }& {\bf Parameter} &{\bf  Value } \\
\hline
\hline
$N_0$&-174\,dBm/Hz&$W$&10\,MHz\\
  \hline
 $\kappa $& $10^{-27}\,\mbox{Watt}\cdot\mbox{s}^3/\mbox{cycle}^3$&$P_{i}^{\max}$&20\,dBm\\
\hline
 $L_i$&$\{737.5,1760,2640,8250\}$&$f_{i}^{\max}$&$10^9$\,cycle/s\\
\hline
$\lambda_i$&$\{0.1, 0.2,\cdots,1.5\}$\,Mbps&$f_{j}^{\max}$&$10^{10}$\,cycle/s\\
\hline
 $d_{i}$&4$\lambda_i$ (bit)&$\epsilon_{i}$&0.01\\
\hline
$\sigma_i^{\rm th}$&4$\lambda_i$ (bit)&$\xi_i^{\rm th}$&0.3\\
\hline
$d_{ji}$&20\,sec& $\epsilon_{ji}$&0.01\\
\hline
  $\sigma_{ji}^{\rm th}$&$4\tilde{R}_{ij}(\infty)$ (bit) &  $\xi_{ji}^{\rm th}$&0.3\\
  \hline
\end{tabular}
\vspace{-1em}
\end {table}
We consider an MEC architecture in which four  servers, each with $9$ CPU cores, are uniformly deployed in a $100\times 100\,\mbox{m}^2$ indoor  area. Additionally, $36$ UEs are randomly distributed, where each UE associates with the nearest server, i.e., $|\mathcal{S}_i|=1,\forall\,i\in\mathcal{U}$. 
For task offloading, we assume that the transmission frequency is 5.8\,GHz with the path loss model $24\log x+20\log 5.8+60$ (dB) \cite{rpt:itu_indoor}, where $x$ in meters is the distance between the transmitter and receiver.   Further,  all wireless channels experience Rayleigh fading with unit variance, with a coherence time of 40\,ms \cite{coherence_time}. Moreover, we consider Poisson  task arrivals. 
The remaining parameters for all UEs  $i\in\mathcal{U}$ and servers $j\in\mathcal{S}$ are listed in Table \ref{Tab: parameters}.\footnote{The four considered processing densities,  $L_i$ (cycle/bit), correspond to the applications of the English Wikipedia main page, 6-queen puzzle, 400-frame video game, and 7-queen puzzle, respectively \cite{JSAC_queue,Mie2010}.}
We first show the tradeoff between  a UE's average power consumption and end-to-end delay in Fig.~\ref{Fig:tradeoff}.  By varying the  non-negative Lyapunov tradeoff parameter $V,$  we obtain the tradeoff curve in which a small $V$ induces a low end-to-end delay at the cost of higher power consumption. In contrast, a large $V$ asymptotically minimizes the power consumption at the expense of a  higher end-to-end latency. 
Given the light processing density and  low task arrival rate, e.g., $L=737.5$\,cycle/bit and $\lambda=0.3$\,Mbps, 
the UE's computation capability is sufficient to execute all tasks locally. In this case, the UE's computation rate is higher than the transmission rate for offloading. Thus, consuming power for task offloading instead of computing locally incurs a worse delay performance. As the processing density increases, or tasks have a higher arrival rate, the UE is  unable to execute tasks  with limited computation capability. Here, consuming power for task offloading achieves a higher task completion rate while the server provides a faster computation capability.  The benefit is more prominent when the processing density is more intense. To summarize, the MEC architecture achieves better power-delay tradeoff performance when  UEs have high-computation tasks or higher task arrival rates.

\begin{figure}[t]
\centering
\includegraphics[width=1\columnwidth]{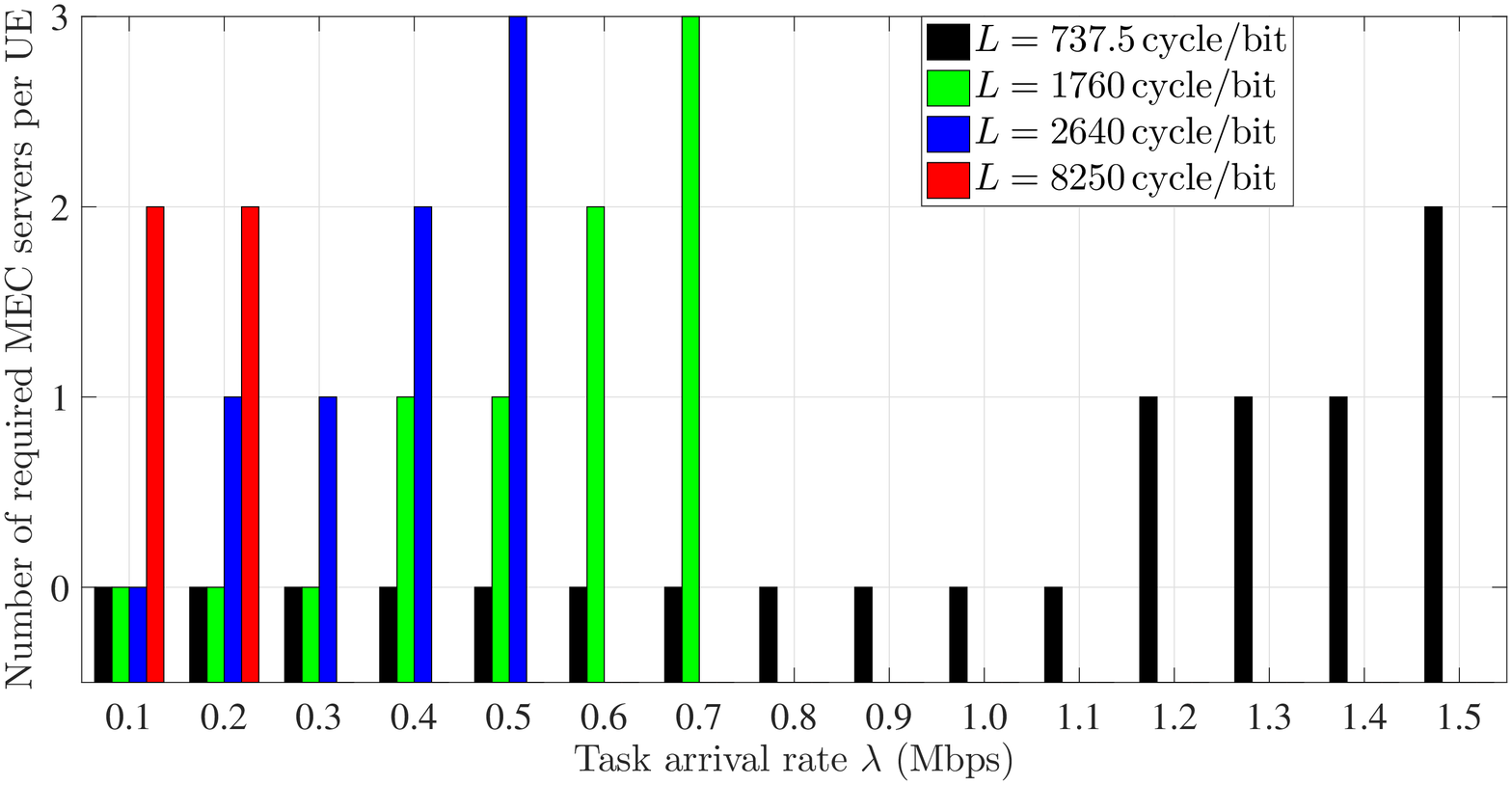}
	\caption{Number of required MEC servers per UE for various processing densities and task arrival rates.}
\label{Fig: Req}
\vspace{-1.5em}
\end{figure}
\begin{figure*}
\centering
\begin{minipage}{0.31\linewidth}
\vspace{-0.9em}
	\includegraphics[width=1\columnwidth]{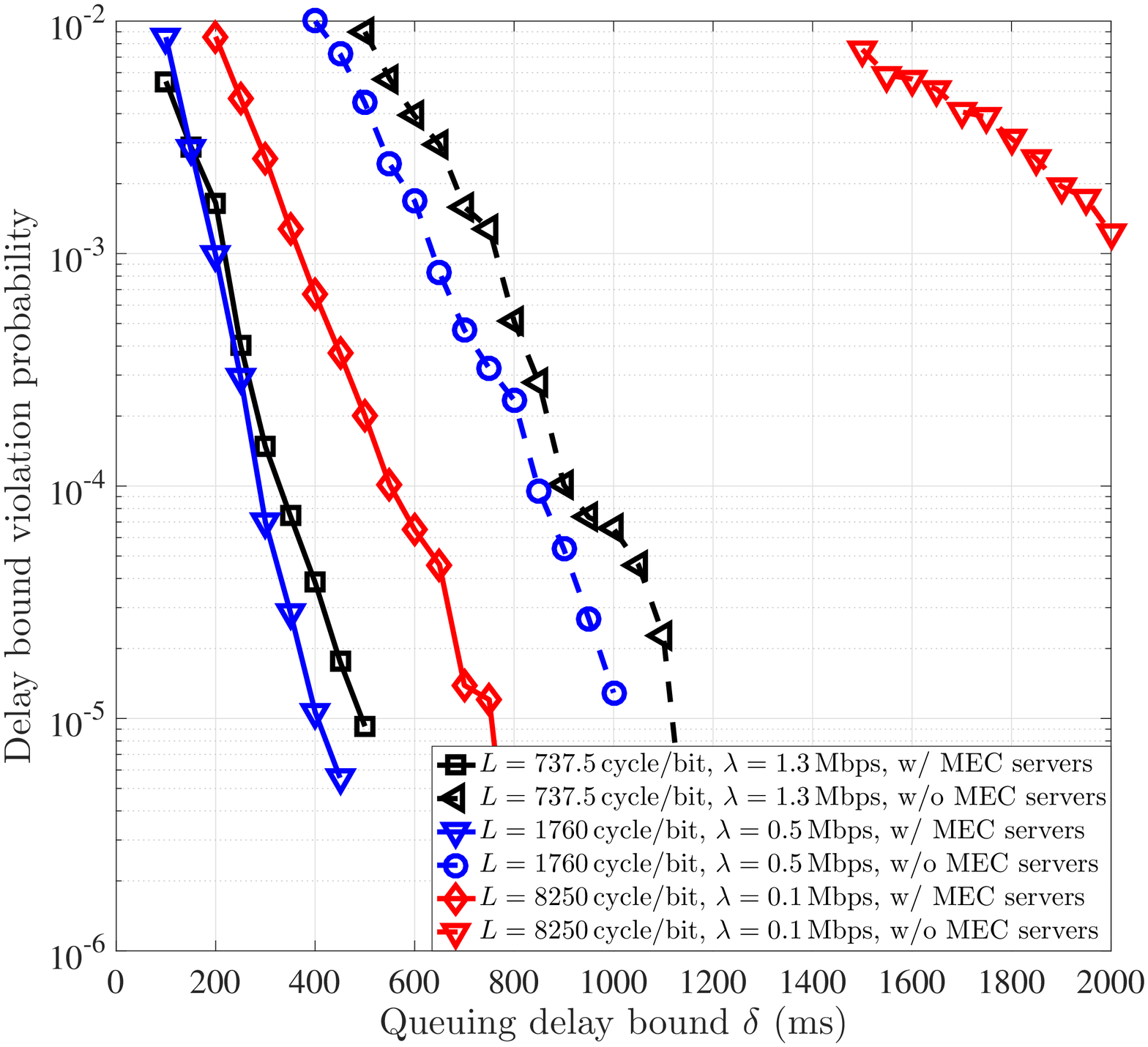}
	\caption{Delay bound violation probability versus queuing delay bound.}
		\label{Fig:reliability}
\end{minipage}
\hspace{0.5em}
\begin{minipage}{0.31\linewidth}
	\includegraphics[width=1\columnwidth]{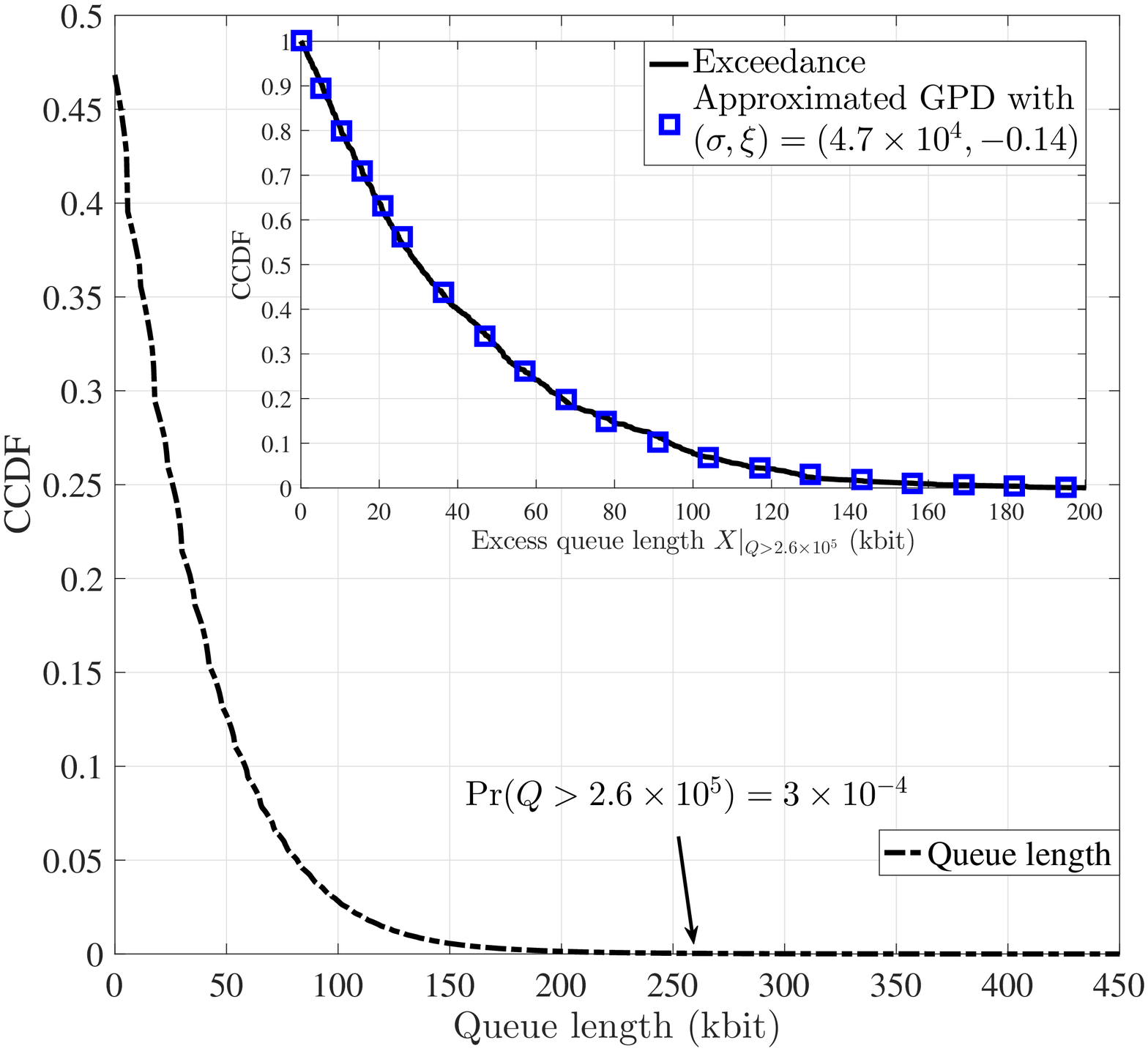}
	\caption{Tail distributions of a UE's task queue length, queue length exceedances over threshold, and the approximated GPD  of exceedances.}
		\label{Fig:tail}
\end{minipage}
\hspace{0.5em}
\begin{minipage}{0.31\linewidth}
\vspace{-0.9em}
	\includegraphics[width=1\columnwidth]{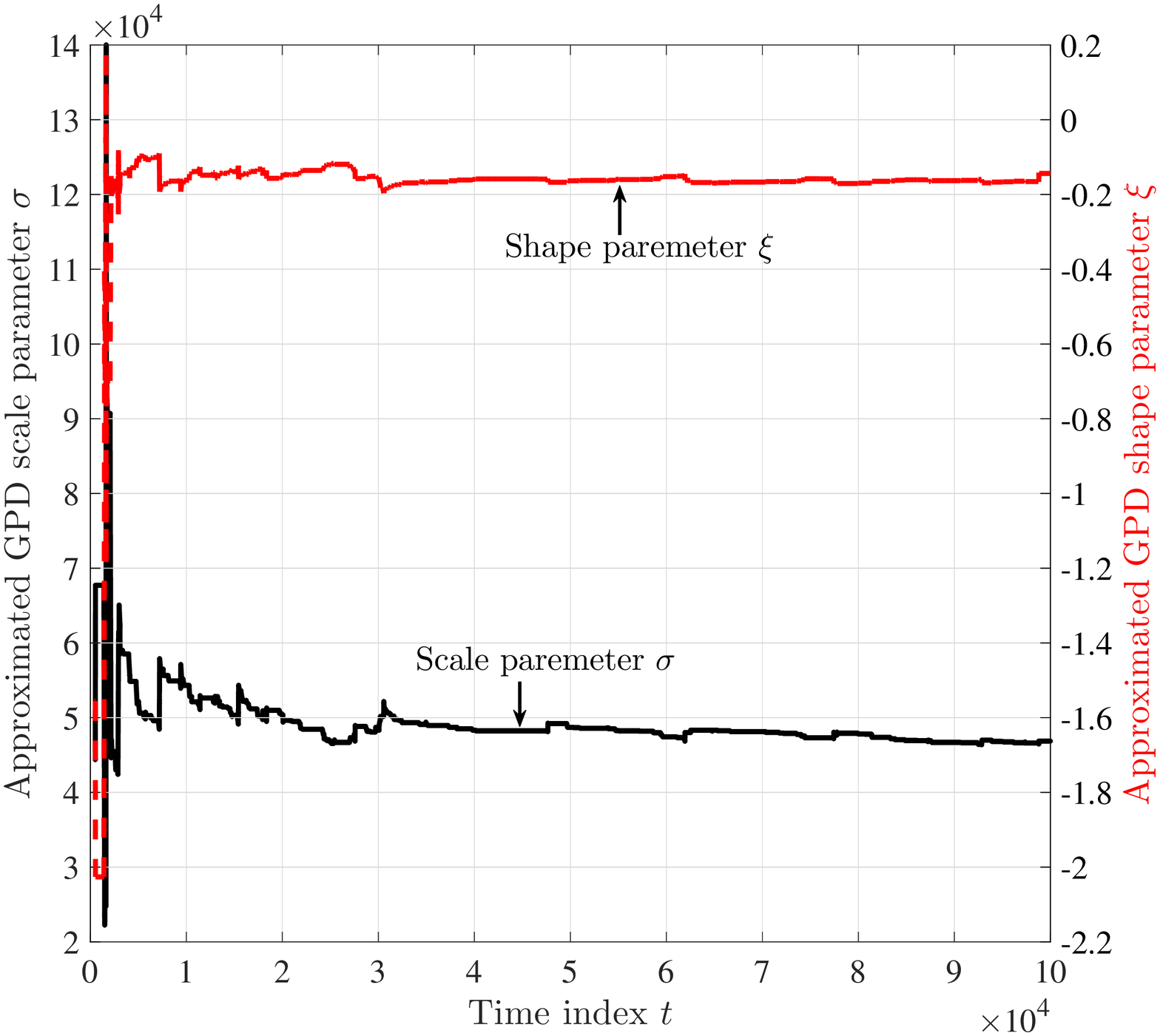}
	\caption{Convergence of the approximated GPD scale and shape parameters of exceedances.}
		\label{Fig:EVT}
\end{minipage}
\vspace{-1.5em}
\end{figure*}

Subsequently, we focus on the performance of the average end-to-end delay and the statistics of the queue length in which the Lyapunov parameter is set to zero  ($V=0$ gives the lowest  average end-to-end delay and queue length \cite{Neely/Stochastic}). 
In Fig.~\ref{Fig:accessed number}, we first show the average end-to-end delay versus the number of  UE accessed servers, i.e., $|\mathcal{S}_i|$, in which each UE $i\in\mathcal{U}$ accesses  $|\mathcal{S}_i|$-nearest  servers.
As discussed in Fig.~\ref{Fig:tradeoff},  the MEC-centric approach outperforms the local computation one without  MEC, i.e., $|\mathcal{S}_i|=0$, in terms of average end-to-end delay when the UE has high computation tasks  or higher task arrival rate. Let us further consider the two settings, 1) $L=737.5$\,cycle/bit and $\lambda=1.3$\,Mbps, and 2) $L=1760$\,cycle/bit and $\lambda=0.5$\,Mbps, in Fig.~\ref{Fig:accessed number}. Although the UE can leverage more computational resources by accessing MEC servers, the multiple UE scenario  incurs severe interference as well as longer waiting times. Therefore, accessing multiple servers does not  decrease the end-to-end delay in these two settings. Nevertheless, if the processing is  intense, the benefit of having access to multiple servers' resources outweighs these  shortcomings.  In Fig.~\ref{Fig:accessed number}, we can find that $|\mathcal{S}_i|=2$ achieves the lowest average end-to-end delay for $L=8250$\,cycles/bit. Based on the results in Fig.~\ref{Fig:accessed number},  Fig.~\ref{Fig: Req} further shows the number of  required MEC servers per UE to achieve the lowest average end-to-end delay given that the task traffic demand with a given processing density can be satisfied (shown by a rectangular bar) in the simulated MEC architecture or local computation scenario.

Further considering the settings (of Fig.~\ref{Fig:accessed number}) in which the MEC-centric approach is superior to the local computation one, we vary the threshold $\delta$ of the probabilistic queuing delay (accounting for both the UE and server) requirement, $\Pr(\mbox{Queuing delay}>\delta)\leq 10^{-2}$, and plot  in Fig.~\ref{Fig:reliability} the exact delay bound violation probability, i.e., reliability. As shown in Figs.~\ref{Fig:accessed number} and \ref{Fig:reliability}, in addition to the average end-to-end delay, the MEC architecture  achieves better reliability performance. In other words, even though there is an extra queuing delay at the server side, offloading tasks to the MEC servers reduces the waiting time for task execution when UEs have high-computation tasks or higher task arrival rates. 
Corresponding to the results in Fig.~\ref{Fig:accessed number}, the reliability enhancement is more prominent while the average end-to-end delay is  reduced, e.g., the case $L=8250$\,cycles/bit.

Finally, let us  investigate the statistics of the queue length.
Due to space limitations, we only plot the UE's task queue length but note that the following illustration is applicable to the offloaded-task queue length at the server. Let us consider $ \lambda_i=1.3$\,Mbps, $L_i=737.5$\,cycle/bit, $d_{i}=2.6\times 10^5$\,bit, and $d_{ji}=0.2$\,sec, $\forall\,i\in\mathcal{U},j\in\mathcal{S}$.
Fig.~\ref{Fig:tail} shows the tail distribution, i.e., CCDF, of the UE's queue length in which we not only ensure the probabilistic constraint on the queue length but also achieve a CCDF value, i.e., $\Pr(Q>2.6\times 10^5)=3\times 10^{-4}$, approaching zero. Applying Theorem \ref{Thm: Pareto} to the conditional excess queue value $X|_{Q>2.6\times 10^5}=Q-2.6\times 10^5$, we  show the tail distributions of the conditional excess queue value and the approximated GPD which coincide with each other.
In Fig.~\ref{Fig:EVT}, we show the convergence of the scale and shape parameters of the approximated GPD. Once convergence is achieved, we can estimate the statistics of the extreme queue length/queuing delay which enables us to proactively deal with the occurrence of extreme events.

\section{Conclusions}\label{Sec: conclude}
In this work, we have studied an URLLC-enabled MEC architecture with multiple users and servers, in which the high-order statistics of latency and reliability are taken into account. To this end, we have imposed a probabilistic constraint on the queue length (or queuing delay) violation and invoked extreme value theory to deal with low-probability (extreme) events.
The studied problem has been cast as a computation and transmit power minimization problem subject to latency requirements and reliability constraints. Subsequently, utilizing Lyapunov stochastic optimization, we have proposed a dynamic latency and reliability-aware  policy for task computation, task offloading, and resource allocation. Numerical results have shown that the considered MEC network achieves a better power-delay tradeoff for intense computation requirements and higher task arrival rates.

\section*{Acknowledgments}
This research was supported by the Academy of Finland project CARMA, the Nokia Bell-Labs project FOGGY, the Nokia Foundation, Emil Aaltosen s\"a\"ati\"o, and the U.S. National Science Foundation under Grants CCF-1420575 and CNS-1702808. The authors are also grateful to Dr.~Chih-Ping Li from Qualcomm Technologies, Inc.~for his very valuable discussions and feedback on this work.

\bibliographystyle{IEEEtran}
\bibliography{ref_fog}
\end{document}